\documentclass[12pt]{article}

\def\slash#1{\setbox0=\hbox{$#1$}#1\hskip-\wd0\hbox to\wd0{\hss\sl/\/\hss}}
\usepackage{epsf, cite}
\usepackage{epsfig}                  
\usepackage[all]{xy}                 
\usepackage{amsfonts}
\usepackage{latexsym}
\usepackage{amsmath}
\usepackage{amssymb}

\usepackage{epsf, cite}

\makeatletter
\renewcommand\section{\@startsection {section}{1}{\z@}%
                                   {-3.5ex \@plus -1ex \@minus -.2ex}
                                   {2.3ex \@plus.2ex}%
                                   {\normalfont\large\bfseries}}
\renewcommand\subsection{\@startsection{subsection}{2}{\z@}%
                                     {-3.25ex\@plus -1ex \@minus -.2ex}%
                                     {1.5ex \@plus .2ex}%
                                     {\normalfont\bfseries}}
\makeatother

\let\non\nonumber

\usepackage{hyperref}

\def\beq{\begin{equation}}
\def\eeq{\end{equation}}
\def    \bea    {\begin{eqnarray}}
\def    \eea    {\end{eqnarray}}

\newcommand{\Ab}{{\bar{A}}}
\newcommand{\Bb}{{\bar{B}}}

\def\ab{{\bar{\alpha}}}

\def\ar{{\bar{a}}}

\renewcommand{\a}{\alpha}
\renewcommand{\b}{\beta} 
\renewcommand{\c}{\sigma}
\def\r{\rho}
\def\ta{\tau}
\def\s{\sigma}
\newcommand{\dl}{\delta}

\renewcommand{\t}{\theta}
\def\g{\gamma}
\def\m{\mu}
\def\n{\nu}
\def\k{\kappa}
\def\l{\lambda}
\renewcommand{\H}{\mathcal{H}}
\def\M{\mathcal G}
\def\L{\mathcal L}
\def\K{\mathcal K}
\def\V{\mathcal V}
\def\R{\mathcal R}
\newcommand{\X}{\mathbb{X}}
\renewcommand{\d}{\partial} 
\newcommand{\G}[3]{\Gamma^{#1}_{\ #2 #3}}

\def\Dh{\hat{D}}
\providecommand{\openone}{\leavevmode\hbox{\small1\kern-3.8pt\normalsize1}}
\def\xt{\tilde{X}}
\def\dt{\tilde{\d}}
\renewcommand{\d}{\partial} 
\renewcommand{\u}[1]{\underline{#1}}

\begin{document}
\pagestyle{plain}
\begin{titlepage}

\begin{center}

\hfill{} \\

\vskip 1cm

{{\Large \bf Connecting T-duality invariant theories} \\

\vskip 1.25cm {Neil B. Copland\footnote{email: neil.copland@vub.ac.be} }
\\
{\vskip 0.2cm
Theoretische Natuurkunde, Vrije Universiteit Brussel,\\
and International Solvay Institutes,\\
Pleinlaan 2,\\
B-1050 Brussels, Belgium\\
}}
\end{center}
\vskip 1 cm

\begin{abstract}
\baselineskip=18pt\

We show that the vanishing of the one-loop beta-functional of the doubled formalism (which describes string theory on a torus fibration in which the fibres are doubled) is the same as the equation of motion of the recently proposed generalised metric formulation of double field theory restricted to this background: both are the vanishing of a generalised Ricci tensor. That this tensor arises in both backgrounds indicates the importance of a new doubled differential geometry for understanding both constructions.

\end{abstract}

\end{titlepage}

\pagestyle{plain}

\baselineskip=19pt

\section{Introduction}

T-duality is a special property of string theory, it requires winding round compact dimensions and so cannot be present in a theory of point particles. Given its importance, there have been many attempts over the years to make it manifest in a string action, but usually this comes at a price. One is usually led to double the number of fields, augmenting the usual co-ordinates, which are dual to momenta, with additional co-ordinates, which are dual to the winding number. In order to maintain the requisite number of degrees of freedom some additional constraint is imposed, and often some other symmetry of the Lagrangian is no longer manifest, such as Lorentz or gauge invariance. 

Recently Hull and Zwiebach proposed double field theory\cite{Hull:2009mi}, taking inspiration from closed string field theory, in which in the presence of toroidal directions the string field must depend on the co-ordinate dual to the winding mode (alternatively this dual co-ordinate can be thought of as the difference of left- and right-moving pieces, where the ordinary co-ordinate is their sum, $\tilde{x}=x_L-x_R$, $x=x_L+x_R$). This theory was formulated to be a genuinely doubled theory, with doubled fields depending fully on the doubled set of co-ordinates. For consistency it was required to include the Kalb-Ramond anti-symmetric 2-form, $b$, and dilaton as well as the metric, $g$, and was constructed to cubic order in perturbations around a background, where the fields were required to obey a constraint arising from the level matching condition of closed string theory; all fields had to be annihilated by the operator $\tilde{\d}_i\d^i$. Although the double field theory possessed a new gauge symmetry to this order, this was not at all manifest in the form of the action and certainly not trivial to show.

It was then shown that on imposition of a stronger form of the constraint, namely that $\tilde{\d}_i\d^i$ should not only annihilate fields, but also products of fields, a background independent action could be written\cite{Hohm:2010jy}. Further, this action took a much more elegant form when written in terms of the `generalised metric' $\H$ \cite{Hohm:2010pp}, which takes the form
\bea\label{Hhhzint}
\H_{AB} &=& \left( \begin{array}{cc}
 h^{-1} & -h^{-1}b\\
 bh^{-1}& h - bh^{-1}b
\end{array}\right)\, ,
\eea
 whereas it had previously been written in terms of ${\cal E}=h+b$. This `generalised metric' is so-called as it also appears in generalised geometry\cite{Gualtieri:2003dx} as well as many other attempts to fashion a T-duality invariant theory. In fact, the action could even be written as an Einstein-Hilbert action for a `generalised Ricci scalar' $\R$, and the equation of motion of the generalised metric $\H$ was the vanishing of a generalised Ricci tensor $\R_{MN}$. 
 
 The strong constraint is so strong it implies that the theory is no longer truly doubled, an $O(d,d)$ rotation can bring it to a frame where the fields depend only on the original co-ordinates $x$ and not the $\tilde{x}$, and it can be shown to be equivalent to the undoubled effective theory of the massless closed string fields. Construction of the full double field theory without the strong constraint remains an open problem. Double field theory can also be written in terms of a double vielbein which makes clear the connection to the early work of Siegel\cite{Siegel:1993th,Siegel:1993xq}.
 
An alternative approach is to take a more worldsheet perspective and look at a sigma model with a doubled target space. In \cite{Hull:2006va,Hull:2004in} Hull formulated a sigma model with a target it space which is a torus fibration (some earlier related work can be found in \cite{Duff:1989tf,Tseytlin:1990nb,Tseytlin:1990va,Maharana:1992my,Schwarz:1993vs}). The torus is then doubled, with the fields all depending only on the base co-ordinates. This formalism as again centred around the generalised metric $\H$. Again the formalism was classically equivalent to the ordinary string sigma model, but now with T-duality made manifest. Quantum equivalence was demonstrated using various methods in \cite{Berman:2007vi,Berman:2007xn,HackettJones:2006bp}. The price of doubling the fibre co-ordinates was to introduce a chirality constraint, the content of which was to say that in a certain frame half the co-ordinates were left moving chiral bosons and the other half were right moving, thus ensuring no excess degrees of freedom. In \cite{Berman:2007xn} this constraint was incorporated into the action, the price of this being that Lorentz invariance was no longer manifest.
 
 The method of \cite{Berman:2007xn} was to perform a background field expansion of this chiral sigma model to one loop and examine the conditions for the vanishing of the beta-functional. Recall that for the standard string sigma model with target space metric $g$ it is precisely this calculation that gives the requirement of vanishing Ricci tensor of the background as the consistency condition following from Weyl invariance, determined via the vanishing of the one-loop beta-functional of $g$\cite{Callan:1985ia}. Including the Kalb-Ramond field and dilaton leads to more involved beta-functionals for each, the vanishing of which is equivalent to the Euler-Lagrange equations following from a particular action; this is then the well-known string effective action. 
 
 The requirement that the beta-functional of $\H$ vanishes in the doubled formalism is the vanishing of a particular tensor, $W$. 
In this paper we demonstrate that in the particular set up of the doubled formalism - a torus fibration with the fields only depending on the base co-ordinates - the generalised Ricci tensor of the double field theory reduces to the tensor $W$. This implies that the double field theory is the effective field theory for the doubled formalism in this set-up. In some sense this is not surprising as both are equivalent to the ordinary string, however, the appearance of the generalised Ricci tensor (which is not just the Ricci tensor of the generalised metric) in both formalisms is indicative of its importance in trying to find a more geometric understanding of the doubled field theory. It also raises the question of whether there is a more general chiral sigma model which leads to the generalised metric formulation of the double field theory as its one-loop effective theory. We will return to these issues in the discussion section.

The structure of this paper is as follows: first we shall briefly review the relevant features of the doubled formalism and the background field expansion before rederiving the one-loop beta-functional. In doing so we shall proceed differently to \cite{Berman:2007xn} in what is ultimately a more satisfying manner, answering some questions that were not fully examined before about the differences in the calculation for ordinary and doubled chiral sigma models. Then we will introduce the double field theory in its generalised metric formulation and hence show the equivalence of the vanishing of the one-loop beta-functional of the doubled formalism to the equations of motion of the doubled field theory in the fibred set-up, including the dilaton terms that were obtained in \cite{Berman:2007yf}. We end with a brief discussion.

\section{Background field expansion in the doubled formalism}

First we introduce the doubled formalism, and then provide a more covariant derivation of the results of \cite{Berman:2007xn} in which the requirements for the vanishing of the one-loop beta-functional were obtained. This is the vanishing of a particular tensor, which we will ultimately compare with double field theory generalised Ricci tensor. Extended details of the calculation of \cite{Berman:2007xn} and related issues can also be found in \cite{Thompson:2010sr}.

\subsection{The doubled formalism}

In order to rewrite the string sigma model in a way in which T-duality is manifest, Hull introduced the doubled formalism\cite{Hull:2004in,Dabholkar:2005ve,Hull:2006va,Hull:2006qs}. On a target space which is locally a $T^n$ bundle the fibre co-ordinates are doubled to $T^{2n}$ with T-duality acting as $O(n,n)$ rotations of this new doubled fibre. The apparent increase in degrees of freedom is compensated by the introduction of a constraint. The Lagrangian, for $d$ base co-ordinates $Y^a$ and $2n$ doubled fibre co-ordinates $\X^A$ is given by\footnote{Here we do not include a topological term, which plays no role here or a possible 1-form connection for the fibration $\mathcal{A}^A(Y)$. We also use the conventions of  \cite{Hull:2006va} and \cite{Berman:2007xn}, in particular we have dropped a factor of $2\pi$ multiplying all Lagrangians.}
\beq
\mathcal{L} = \frac{1}{4} \H_{AB}(Y)d\X^A\wedge\ast d\X^B  +
\mathcal{L}(Y) \, ,
\eeq
where $\mathcal{L}(Y)$ is the standard string sigma model Lagrangian on the
base and $\H(Y)$ is a metric on the fibre.

One can choose a frame where $\H$ has the `generalised metric' form
\bea\label{H}
\H_{AB}(Y) &=& \left( \begin{array}{cc}
 h^{-1} & -h^{-1}b\\
 bh^{-1}& h - bh^{-1}b
\end{array}\right)\, ,
\eea
$h$ and $b$ are the target space metric and $B$-field on the fibre of
the undoubled fibre. In this frame $\X^A=(X^i,\xt_i)$ 
with $\{\xt_i\}$ the coordinates on the T-dual torus. Indices $A,B,\ldots$ are raised and lowered with
\bea  L_{AB} = \left( \begin{array}{cc}
0& \openone\\
\openone & 0
\end{array}\right)\, ,
\eea
the $O(n,n)$ invariant metric as we have $\H^{-1}=L^{-1}\H L^{-1}$.
This is the same as the matrix (\ref{Hhhzint}) except for the allowed co-ordinate dependence, and we have chosen conventions here to agree with \cite{Hohm:2010pp} which means that $\H$ is the inverse of the conventions \cite{Berman:2007xn}. As this theory with $\H$ inverted is T-dual to the original theory, and both formalisms are by construction T-duality invariant it is trivial to change the conventions, and we will not deal with explicit expressions containing $b$ and $h$ anyway.

The constraint which makes sure we still have the requisite number of degrees of freedom after doubling the fibre is
\beq
\label{eqConstraint}
d\X^A = L^{AB}\H_{BC}\ast d\X^C ,
\eeq
Introducing a vielbein $\V^{A}_{\ \Ab}$ one can shift to the chiral frame where 
\bea
\H_{\Ab\Bb}(y) = \left( \begin{array}{cc}
\openone &0\\
0 & \openone
\end{array}\right), &  L_{\Ab\Bb} = \left( \begin{array}{cc}
\openone &0\\
0 & -\openone
\end{array}\right).
\eea
In this frame the constraint (\ref{eqConstraint}) is a chirality
constraint ensuring that half the $\X^\Ab$ are chiral Bosons and the other half
are anti-chiral Bosons.
In \cite{Berman:2007xn} it was shown that using the PST procedure\cite{Pasti:1996vs} to impose the chirality constraint in this frame leads to the action

\bea\label{Daction}
S= \frac{1}{2} \int d^2\sigma\left[ -\M_{\a\b} \partial_1 X^\a \partial_1 X^\b + \L_{\a\b} \partial_1 X^\a \partial_0 X^\b + \K_{\a\b} \partial_0 X^\a \partial_0 X^\b\right]\,   ,
\eea
where $X^\a = (\X^A, Y^a)=(X^i, \tilde{X}_j, Y^a)$ and
\beq
\M=
\left(
\begin{array}{cc}
\H  &   0 \\
 0 &    g 
\end{array}
\right),
\L=
\left(
\begin{array}{cc}
L &   0 \\
 0 &    0
\end{array}
\right),
\K=
\left(
\begin{array}{cc}
0  &   0 \\
 0 &    g
\end{array}
\right).
\eeq
The equation of motion for the fibre co-ordinates is 
\bea
\d_1 \left( \H \d_1 \X \right) = L \d_1\d_0 \X,
\eea
which integrates to give the constraint (\ref{eqConstraint})\footnote{Where we use the gauge invariance of the action under $\X^A\rightarrow \X^A+f(\tau)$ to remove an integration function of $\ta$.}.  

\section{The background field expansion}

The background field expansion\cite{AlvarezGaume:1981hn,Honerkamp:1971sh,Braaten:1985is} allows the study the UV divergences perturbatively by expanding quantum fluctuations around a classical background which solves the equations of motion. The expansion is chosen carefully to preserve the covariance of the original action. Consistency of the ordinary string sigma model requires Weyl invariance, and at one loop this is equivalent to the vanishing of the beta-functional which famously requires that the Ricci tensor is zero\cite{Callan:1985ia}. Since the constraint (\ref{eqConstraint}) is now incorporated into the action (\ref{Daction}), following \cite{Berman:2007xn} we can find the requirement for the vanishing of the beta-functional in the doubled formalism.

In \cite{Berman:2007xn} the algorithmic method of calculating the background field expansion developed in \cite{Mukhi:1985vy} was used (the details can be found in either). In the following expressions $X^\a$ is the background field and $\xi^\a$ represents the quantum fluctuation. The first-order term in the expansion in $\xi$  is proportional to the equations of motion of $X$ and vanishes, for a one-loop calculation we need the second-order terms in the fluctuations. These are given by
\bea\label{fulag}
2{\mathcal L}_{(2)}&=&-\M_{\a\b}D_1\xi^\a D_1\xi^\b+\L_{\a\b}D_0\xi^\a D_1\xi^\b+\K_{\a\b}D_0\xi^\a D_0\xi^\b\non\\
&&-R_{\g\a\b\dl}\xi^\a\xi^\b\d_1X^\g\d_1X^\dl+\L_{\a\b;\g}\xi^\g(D_0\xi^\a\d_1X^\b+\d_0X^\a D_1\xi^\b)\non\\
&& +\frac{1}{2}D_\a D_\b\L_{\g\dl}\xi^\a\xi^\b\d_0X^{\g}\d_1X^\dl
+\frac{1}{2}\left( \L_{\g\s}R^{\s}_{\ \a\b\dl} + \L_{\dl\s}R^{\s}_{\ \a\b\g}\right)\xi^\a\xi^\b\d_0X^\g\d_1X^\dl\non\\
&&+2\K_{\a\b;\g}\xi^\g D_0\xi^\a\d_0X^\b\non\\
&&  +\frac{1}{2}D_\a D_\b\K_{\g\dl}\xi^\a\xi^\b\d_0X^{\g}\d_0X^\dl
+\K_{\g\s}R^{\s}_{\ \a\b\dl}\xi^\a\xi^\b\d_0X^\g\d_0X^\dl \, ,
\eea
and the effective action at second order is given by the integral of this Lagrangian.

\subsection{Contracting maintaining covariance}

The next step to obtaining the beta-functional is to Wick contract using propagators for the fluctuations. Details of how to obtain the propagators are given in the appendix of \cite{Berman:2007xn}, in the chiral frame the fluctuation kinetic terms are those of chiral bosons in flat space and we can obtain the propagators using techniques of \cite{Tseytlin:1990nb,Tseytlin:1990va}. In moving the indices on fluctuations in our Lagrangian to the chiral frame the forfeit is that we must introduce derivatives acting on vielbeins as $\d_\mu \xi^A=\V^A_{\ \Ab}\d_\mu \xi^\Ab+\d_\mu \V^A_{\ \Ab} \xi^\Ab$. Thus we can use the contractions as if the indices are in the metric frame if we replace the connection in covariant derivatives of fluctuations by the spin connection, 
\beq
D_\mu\xi^A=\d_\mu\xi^A+\G{A}{\mu}{B}\xi^{B}\rightarrow\d_\mu\xi^A+\G{A}{\mu}{B}\xi^{B}+\d_\mu\V^A_{\ \Ab}\V^{\Ab}_{B}\xi^B=\d_\mu\xi^A+A^{A}_{\ \mu B}\xi^{B}\, ,
\eeq
where we take this as the definition of $A^{A}_{\ \mu B}$. We use the notation $A^{A}_{\ \mu B}$ because in the case of the non-doubled string sigma model there is a general argument that $A^{A}_{\ \mu B}$ cannot contribute to the one-loop divergence as it transforms as an $SO(n)$ gauge potential\cite{AlvarezGaume:1981hn}. For now we will assume these terms will not contribute and drop them, but we shall return to the issue later.

The contraction of two fluctuations is given by
\bea
\langle \xi^{\a}(z) \xi^{\b}(z) \rangle 
&=& \Delta_0\M^{\a\b}  +\theta \L^{\a\b},
\eea
where $\Delta_0$ is the propagator of an ordinary boson (terms proportional to $\theta$ give a possible Lorentz anomaly, this was shown to vanish in  \cite{Berman:2007xn} and we do reconsider them  here). Terms in the effective action containing two fluctuation fields $\xi$ without derivatives acting on them can be Wick contracted with the above propagator eliminating the fluctuations and will contribute at one loop, we write these single contraction terms as $\L_{(2s)}$. To find all possible one loop terms should consider the exponential of the effective action, and examining the structure of possible terms there will also be contributions from terms at second order in the exponential which contain four fluctuations, two of which have derivatives. To calculate their contribution we need the following four-fluctuation contractions \footnote{In these expressions we correct a factor of $-1/2$ compared to the presentation in \cite{Berman:2007xn}.}
\bea
 \langle \xi^\g \d_1 \xi^\a \d_1 \xi^\r \xi^\ta \rangle &=& -\Delta_0\left(\M^{\a[\ta} \M^{\r]\g} -\L^{\a[\ta} \L^{\r]\g}\right),\\
\langle \xi^\g \d_1 \xi^\a \d_0 \xi^\r \xi^\ta \rangle &=& \Delta_0\left(\M^{\a[\ta} \L^{\r]\g}+\L^{\a[\ta} \M^{\r]\g}\right)+  \t \L^{\a[\ta} \L^{\r]\g},   \\
\langle \xi^\g \d_0 \xi^\a \d_0 \xi^\r \xi^\ta \rangle &=& \Delta_0\left(\M^{\a[\ta} \M^{\r]\g} +3\L^{\a[\ta} \L^{\r]\g}\right) +  \t \left( \M^{\a[\ta} \L^{\r]\g}+\L^{\a[\ta} \M^{\r]\g}\right),   
\eea
(assuming we have introduced the connection $A$, otherwise the indices should be in the tangent space). These double contraction terms (which we will write as $\L_{(2d)}$, we will also write $\L_{(2db)}$ for the base part of this etc.) come a factor of $1/4$ down on those in $\L_{(2s)}$ due to the half in the exponential of the action and another half from squaring the overall factor of a half that comes with $\L_{(2s)}$ that we have chosen to write on the right hand side to make expressions clearer. 

In \cite{Berman:2007xn} at this stage multiples of the equation of motion were added to simplify matters  and the covariance of the expressions was lost. We will proceed differently, maintaining the covariance of the background field expansion as long as possible and not utilising the equation of motion unless strictly necessary.

The terms with a single contraction are
\bea\label{SingleCon}
2\L_{(2s)}/\Delta_0&=&R_{\g\dl}\d_1X^\g\d_1X^\dl +\frac{1}{2}\left(\M^{\a\b}D_\a D_\b\L_{\g\dl}-\L_{\g\s}R^{\s}_{\ \dl} -\L_{\dl\s}R^{\s}_{\ \g}\right)\d_0X^{\g}\d_1X^\dl\label{singleRL}\qquad\\
&&+\frac{1}{2}\left(\M^{\a\b}D_\a D_\b\K_{\g\dl}- 2\K_{\g\s}R^{\s}_{\ \dl} \right)\d_0X^{\g}\d_0X^\dl\label{singleK}\, .
\eea
The double contraction terms which also contribute the the divergence enter at second-order in the expansion of the exponential of the effective action and can be found by squaring the terms in the effective action with Lagrangian (\ref{fulag}) which have two fluctuations, one of which has a propagator, i.e the terms
\beq\label{3xidxi}
2\sqrt{2\L_{(2d)}}=\L_{\a\b;\g}\xi^\g(D_0\xi^\a\d_1X^\b+\d_0X^\a D_1\xi^\b)+2\K_{\a\b;\g}\xi^\g D_0\xi^\a\d_0X^\b,
\eeq
and this leads to 6 possible terms in $\L_{(2d)}$.

\subsection{Base terms}

Recalling that $\H^{AB}\d_g\H_{AB}=0$ and we find from the single contraction terms (\ref{SingleCon})
\bea
2\L_{(2sb)}/\Delta_0&=&R_{gd}\d_1X^g\d_1X^d\\
&&+\frac{1}{4}(\H^{-1}\d_g\H L^{-1}\d_d\H)\d_0X^g\d_1X^d\\
&&+(\frac{1}{4}\d_g\H^{-1}\d_d\H-R_{gd})\d_0X^g\d_0X^d\, ,
\eea
with the middle line vanishing due to properties of $\H$ and $L$. Note that $R_{gd}$ is the bas component of the doubled Ricci tensor, $R(\H)$, as opposed to the Ricci tensor of $g_{gd}$, which we will denote $\hat{R}$. The double contraction terms then give
\bea
&&\L_{\a b;\g}\L_{\r s;\ta}\langle\xi^\g \d_0\xi^\a \d_0\xi^\r\xi^\ta\rangle\d_1X^b\d_1X^s=-\frac{1}{8}(\d_g\H^{-1}\d_d\H)\d_1X^g\d_1X^d\, ,\\
&&2\L_{\a b;\g}\L_{\r s;\ta}\langle\xi^\g \d_0\xi^\a \d_1\xi^\r\xi^\ta\rangle\d_1X^b\d_0X^s=0\, ,\\
&&\L_{\a b;\g}\L_{\r s;\ta}\langle\xi^\g \d_1\xi^\a \d_1\xi^\r\xi^\ta\rangle\d_0X^b\d_0X^s=-\frac{1}{8}(\d_g\H^{-1}\d_d\H)\d_0X^g\d_0X^d\, ,\\
&&4\L_{\a b;\g}\K_{\r s;\ta}\langle\xi^\g \d_0\xi^\a \d_0\xi^\r\xi^\ta\rangle\d_1X^b\d_0X^s=0\, ,\\
&&4\L_{\a b;\g}\K_{\r s;\ta}\langle\xi^\g \d_1\xi^\a \d_0\xi^\r\xi^\ta\rangle\d_0X^b\d_0X^s=0\, ,\\
&&4\K_{\a b;\g}\K_{\r s;\ta}\langle\xi^\g \d_0\xi^\a \d_0\xi^\r\xi^\ta\rangle\d_0X^b\d_0X^s=0\, .\
\eea
Thus the grand total on the base is
\beq
\L_{(2b)}=\frac{1}{2}W_{gd}\d_\mu X^g\d^\mu X^d\Delta_0
\eeq
where
\beq\label{BaseWeyl}
W_{gd}=-\hat{R}_{gd}-\frac{1}{8}\d_g\H^{-1}\d_d\H=-(R_{gd}-\frac{1}{8}\d_g\H^{-1}\d_d\H)\, .
\eeq
This agrees with result of \cite{Berman:2007xn}. That result was calculated without dropping all terms involving the connection $A$, whereas those terms were dropped here. We  conclude that they do not contribute to the divergence on the base, just as for the standard sigma model. Note that in \cite{Berman:2007xn} it was stated that `vielbein terms' did contribute in general, unlike the ordinary string case, but the connection $A$ is a combination of an ordinary Christoffel symbol and the vielbein piece which has derivative acting on a vielbein, so the statements are compatible.

\subsection{Fibre terms}

First we look at the single contraction terms. The Ricci tensor $R$, the result we expect, is sitting there as the first term of (\ref{SingleCon}). We also find
\beq\label{DDLetc}
\frac{1}{2}\left(\M^{\a\b}D_\a D_\b\L_{\g\dl}-\L_{\g\s}R^{\s}_{\ \dl} -\L_{\dl\s}R^{\s}_{\ \g}\right)\d_0X^{\g}\d_1X^\dl=
\frac{1}{4}(\d_a\H)L^{-1}(\d^a\H)_{GD}\d_0X^G\d_1X^D.
\eeq
with the first term cancelling the second and third terms and providing the extra piece.
Though not obvious, the $\K$ terms also contain a contribution on the fibre (one should remember not to restrict indices to the base or fibre before acting with covariant derivatives) given by
\bea\label{DDK}
\frac{1}{2}\M^{\a\b}D_\a D_\b\K_{\g\dl}\d_0X^{\g}\d_0X^\dl=
\frac{1}{4}(\d_a\H)\H^{-1}(\d^a\H)_{GD}\d_0X^G\d_0X^D.
\eea
The six double contraction terms are
\bea\label{DubConFib}
\L_{\a B;\g}\L_{\r S;\ta}\langle\xi^\g \d_0\xi^\a \d_0\xi^\r\xi^\ta\rangle\d_1X^B\d_1X^S&=&-1/32(\d_a\H)\H^{-1}(\d^a\H)_{GD}\d_1X^G\d_1X^D,\quad\ \ \ \\
2\L_{\a B;\g}\L_{\r S;\ta}\langle\xi^\g \d_0\xi^\a \d_1\xi^\r\xi^\ta\rangle\d_1X^B\d_0X^S&=&1/{16}(\d_a\H)L^{-1}(\d^a\H)_{GD}\d_1X^G\d_0X^D,\\
\L_{\a B;\g}\L_{\r S;\ta}\langle\xi^\g \d_1\xi^\a \d_1\xi^\r\xi^\ta\rangle\d_0X^B\d_0X^S&=&-1/32(\d_a\H)\H^{-1}(\d^a\H)_{GD}\d_0X^G\d_0X^D,\\
4\L_{\a B;\g}\K_{\r S;\ta}\langle\xi^\g \d_0\xi^\a \d_0\xi^\r\xi^\ta\rangle\d_1X^B\d_0X^S&=&-1/8(\d_a\H)L^{-1}(\d^a\H)_{GD}\d_1X^G\d_0X^D,\\
4\L_{\a B;\g}\K_{\r S;\ta}\langle\xi^\g \d_1\xi^\a \d_0\xi^\r\xi^\ta\rangle\d_0X^B\d_0X^S&=&-1/8(\d_a\H)\H^{-1}(\d^a\H)_{GD}\d_0X^G\d_0X^D,\\
4\K_{\a B;\g}\K_{\r S;\ta}\langle\xi^\g \d_0\xi^\a \d_0\xi^\r\xi^\ta\rangle\d_0X^B\d_0X^S&=&-1/8(\d_a\H)\H^{-1}(\d^a\H)_{GD}\d_0X^G\d_0X^D.
\eea
Now the equation of motion is
\beq
D_{1}(\M_{\a\b}\d_1 X^{\b})=\L_{\a\b}\d_1\d_0X^\b+\Dh_{0}(\K_{\a\b}\d_0X^{\b})\, ,
\eeq
where as always hatted quantities are constructed from the base metric $g_{ab}$ only. We have seen the fibre part of this before as the constraint (\ref{eqConstraint}), and we could use this here to cancel terms with different worldsheet index structure off against each other; the single contraction terms other than the Ricci tensor (the result we want) would cancel each other, and we can apply it to the six double-contraction terms beginning at (\ref{DubConFib}), but we find a non-zero remainder  of
\beq
\L_{(2df)}=-\frac{1}{16}(\d_a\H)\H^{-1}(\d^a\H)_{GD}\d_0X^G\d_0X^D.
\eeq
This difference from the result of \cite{Berman:2007xn} tells us that we are not allowed to disregard the connection pieces on the fibre. It now carries doubled $SO(n,n)$ doubled indices, whereas on the base it still transforms as an $SO(d)$ connection and cannot contribute by the arguments of \cite{AlvarezGaume:1981hn}. We will now confirm that including these connection terms gives the correct answer.

\subsection{Connection terms}

Recall that the connection is given by
\beq
A^{\a}_{\m\, \b}=\d_{\m}\V^{\a}_{\ab}\V^{\ab}_{\b}+\G{\a}{\b}{\g}\d_{\m}X^{\g}.
\eeq
We are only interested in terms on the fibre, and since the vielbein is independent of $X^A$, this means the vielbein derivative piece cannot contribute. Further we can split $A$ into its diagonal and off-diagonal parts
\bea
A^{A}_{\m\, B}&=&\d_{\m}\V^{A}_{\Ab}\V^{\Ab}_{B}+\G{A}{B}{g}\d_{\m}X^{g}\, ,\\
A^{A}_{\m\, b}&=&\G{A}{b}{G}\d_{\m}X^{G}\, ,\\
A^{a}_{\m\, b}&=&\d_{\m}\V^{a}_{\ar}\V^{\ar}_{b}+\G{a}{b}{g}\d_{\m}X^{g}\, ,\\
A^{a}_{\m\, B}&=&\G{a}{B}{G}\d_{\m}X^{G}\, ,
\eea
and we see only the off-diagonal pieces will contribute to the fibre divergence. When we examine the connection terms we left out before, once again there will be single and double contraction terms. There are 5 single contraction terms, 2 of which are linear in $A$.
\bea\label{ConSing}
\M_{\a\b}A^{\a}_{1\, \g}A^{\b}_{1\, \dl}\M^{\g\dl}&=&-\frac{1}{2}(\d_a\H)\H^{-1}(\d^a\H)_{GD}\d_1X^D\d_1X^G,\\
\L_{\a\b}A^{\a}_{0\, \g}A^{\b}_{1\, \dl}\M^{\g\dl}&=&\frac{1}{4}(\d_a\H)\L^{-1}(\d^a\H)_{GD}\d_1X^D\d_0X^G,\\
\K_{\a\b}A^{\a}_{0\, \g}A^{\b}_{0\, \dl}\M^{\g\dl}&=&\frac{1}{4}(\d_a\H)\H^{-1}(\d^a\H)_{GD}\d_0X^D\d_0X^G,\\
2\K_{\a\b;\g}A^{\a}_{0\, \dl}\M^{\g\dl}\d_0 X^\b&=&-\frac{1}{2}(\d_a\H)\H^{-1}(\d^a\H)_{GD}\d_0X^D\d_0X^G,\\
\L_{\a\b;\g}\left(A^{\a}_{0\, \dl}\d_1 X^\b+A^{\a}_{1\, \dl}\d_0 X^\b\right)\M^{\g\dl}&=&-\frac{1}{2}(\d_a\H)\L^{-1}(\d^a\H)_{GD}\d_1X^D\d_0X^G.
\eea

In addition there are 4 terms in the action of the form $\sim A\xi\d\xi$. Previously in (\ref{3xidxi}) we had three such terms leading to six second-order two-contraction terms which did not cancel, so it looks like we have 28 more terms. In fact on closer examination 3 of the new $\sim A\xi\d\xi$ connection pieces cancel with the 3 original terms, leaving only one term, which squares to give only one second-order term (thus the 6 terms of (\ref{DubConFib}) are eliminated). To see this let us examine one of the new terms restricted to the fibre
\beq
2\K_{\a\b}A^{\a}_{0\, \g}\xi^\g\d_0\xi^\b|_{fibre}=2\K_{ad}\G{d}{B}{G}\xi^G\d_0\xi^a\d_0X^B=-2\K_{\a\b;\g}\xi^\g \d_0\xi^\a\d_0X^\b|_{fibre}
\eeq
where we recognise the this as cancelling the final term of (\ref{3xidxi}) (we have dropped a term which does not contribute to contractions as it is symmetric in $(\a\g)$).

Thus the only remaining second order term is 
\beq
\M_{\a\b}A^{\a}_{1\, \g}\M_{\r\s}A^{\r}_{1\, \ta} \langle \xi^\g \d_1 \xi^\b \d_1 \xi^\s \xi^\ta \rangle=\frac{1}{2}(\d_a\H)\H^{-1}(\d^a\H)_{GD}\d_1X^D\d_1X^G
\eeq
which cancels with the first line of (\ref{ConSing}) (exactly how it would work in the undoubled case). The additional four terms of (\ref{ConSing}) cancel (\ref{DDLetc}) and (\ref{DDK}) without the use of the constraint. Thus we have rederived the result of \cite{Berman:2007xn} that
\beq
\L_{(2f)}=-\frac{1}{2}W_{GD}\d_1X^G\d_1X^D\Delta_0
\eeq
where $W_{GD}=-R_{GD}$.

\subsection{Doubled renormalisation}
The result is that the the Weyl divergence is given by
\beq
S_{Weyl}=\frac{1}{2}\int d^2 \c \bigl[ -W_{GD} \d_1 \X^G\d_1 \X^D + W_{gd}\d_\mu Y^g \d^\mu Y^d \bigr]\Delta_0 \, .
\eeq
One may now proceed directly to regularise and renormalise the divergences coming from $\Delta_0$ in the standard fashion. The following beta-functionals are obtained for the couplings in the action
\bea
\b^\M_{\a\b} = -\left(\begin{array}{cc}
W_{AB} & 0 \\ 0 & W_{ab}
\end{array}\right)\, , & \b^\K_{\a\b} = -\left(\begin{array}{cc}
0 & 0 \\ 0 & W_{ab} 
\end{array}\right)\, , & \b^\L_{\a\b} = 0 \, .
\eea
 Demanding the vanishing of these beta-functions gives the background field equations, the vanishing of $W_{AB}$ and $W_{ab}$. Recall that while $W_{AB}=-R_{AB}$, 
\beq
W_{gd}=-(R_{gd}-\frac{1}{8}\d_g\H^{-1}\d_d\H)\, .
\eeq
In \cite{Berman:2007xn} it was shown that the vanishing of $W$ reproduces the background field equations of the standard string in this set-up. Although this calculation includes only a metric like object, it is equivalent to the string background field equations including the anti-symmetric Kalb-Ramond field $b$ as well.

\section{Double field theory}

Double field theory is a string field theory inspired approach to describing the massless sting modes on a torus. It was introduced by Hull and Zwiebach in \cite{Hull:2009mi} and developed in \cite{Hohm:2010jy,Hohm:2010pp,Hull:2009zb}, later also with Hohm. Since in string field theory on a torus it is necessary to treat the momentum and winding democratically, it is not surprising that a T-duality invariant theory with the torus co-ordinates and extra co-ordinates dual to the winding emerges. As the formalism developed it was first written in terms of ${\cal E}=h+b$ the sum of the metric and B-field, however more recently it has been recast in terms of the generalised metric 
\bea\label{Hhhz}
\H_{MN} &=& \left( \begin{array}{cc}
 h^{-1} & -h^{-1}b\\
 bh^{-1}& h - bh^{-1}b
\end{array}\right)\, .
\eea
where the doubled co-ordinates are $X^M=(\tilde{x}_m,x^m)$ and the original co-ordinates $x_i$ can be compact or not.

In \cite{Hohm:2010pp}  the action of double field theory was cast as an Einstein-Hilbert term for a generalised Ricci-like scalar $\R$ which is a function of the generalised metric $\H$ and the doubled dilaton, $d$, which is defined by the relation
\beq
e^{-2d}=\sqrt{h}e^{-2\phi}.
\eeq
The action was
\beq\label{Raction}
S=  \int dx \, d\tilde x  \, e^{-2d} \, {\cal R}\;.
\eeq
for the generalised Ricci scalar
\beq
 \label{simplerformR}
 \begin{split}
  {\cal R} \ =  &~~~4\,{\cal H}^{MN}\partial_{M}\partial_{N}d
  -\partial_{M}\partial_{N}{\cal H}^{MN} \\[1.2ex]
   & -4\,{\cal H}^{MN}\partial_{M}d\,\partial_{N}d
   + 4 \partial_M {\cal H}^{MN}  \,\partial_Nd\;,\\[1.0ex]
    ~&+\frac{1}{8}\,{\cal H}^{MN}\partial_{M}{\cal H}^{KL}\,
  \partial_{N}{\cal H}_{KL}-\frac{1}{2}{\cal H}^{MN}\partial_{M}{\cal H}^{KL}\,
  \partial_{K}{\cal H}_{NL}\;.
 \end{split}
 \eeq
The dilaton equation of motion is the vanishing of $\R$, and the variation of the action with respect to $\H^{AB}$  is proportional to $\K_{AB}\delta \H^{AB}$, where 
\bea
\label{kis}
\begin{split}  
{\cal K} _{MN}=
~&~
\frac{1}{8}\, \partial_{M}{\cal H}^{KL}
  \,\partial_{N}{\cal H}_{KL}
  -{1\over 4} (\partial_L - 2 (\partial_L d) ) 
  ({\cal H}^{LK} \partial_K {\cal H}_{MN})
  +2 \,\partial_{M}\partial_N d\,  
    \\[1.0ex]
  & \hskip-10pt -\frac{1}{2} \partial_{(M}{\cal H}^{KL}\,\partial_{L}
  {\cal H}_{N)K}
  + {1\over 2} (\partial_L - 2 (\partial_L d) )  \bigl({\cal H}^{KL} \partial_{(M}
   {\cal H}_{N)K}
  + {\cal H}^K{}_{(M}  \partial_K {\cal H}^L{}_{N)}  \bigr) \,.
   \end{split}
\eea
However, the field equation is not simply the vanishing of $\K_{AB}$, as the variation of $\H$ should preserve its coset form. In other words, since the original field satisfies $\H^{AB}L_{BC}\H^{CD}=L^{AD}$,   the field after variation $\H'=\H+\delta\H$ must satisfy the same relation. This constrains the form of the variation and thus the field equation is the vanishing of\footnote{Or indeed in the notation of \cite{Jeon:2010rw} this could also be written $\R=P\K\bar{P}+\bar{P}\K P$ as the projector appears naturally here.}
\bea
\R_{MN}&=&\frac{1}{4}\left(\delta^{\ P}_{M}-\H^{\ P}_{M}\right)\K_{PQ}\left(\delta^{Q}_{\ N}+\H^{Q}_{\ N}\right)+\frac{1}{4}\left(\delta^{\ P}_{M}+\H^{\ P}_{M}\right)\K_{PQ}\left(\delta^{Q}_{\ N}-\H^{Q}_{\ N}\right)\quad\\
&=&\frac{1}{2}\left(\K_{MN}-\H^{\ P}_{M}\K_{PQ}\H^{Q}_{\ N}\right)\, .
\eea
This gives the `generalised Ricci tensor' that we can compare with the result of the background field expansion of the sigma model (note that clearly from the projection structure $\H^{MN}\R_{MN}=0$, not $\R$). 

Note that we can observe that $W_{MN}$ has the right form in terms of the projections $(1\pm\H)/2$, as in \cite{Berman:2007xn} it observed that using the constraint (\ref{eqConstraint}) we could rewrite 
\beq\label{Wshare}
-\frac{1}{2}W_{MN}\d_1\X^M\d_1X^N\rightarrow\frac{1}{4}W_{MN}\d_\mu\X^M\d^\mu \X^N
\eeq
precisely because $W_{MN}=(W_{MN}-\H^{\ P}_{M}W_{PQ}\H^{Q}_{\ N})/2$.

The action (\ref{Raction}) is invariant under doubled gauge transformations which includes diffeomorphisms of $h$ and ordinary gauge transformations of $b$. The gauge transformations act linearly on $\H$ (unlike their action on ${\cal E}$) which leads to the notion of a generalised Lie derivative whose commutator also introduces a modified Courant bracket for the doubled fields.

\section{Reducing the generalised Ricci tensor}

The double field theory is defined on a much more general manifold that the doubled formalism, so in order to relate the tensors $\R_{MN}$ and $W_{MN}$ we should reduce the double field theory tensor on the appropriate fibred manifold. As we have not been dealing with any global issues it is sufficient for our purposes to reduces on a trivial bundle, the main points being that we the fields only depend on the base co-ordinates, and we undouble the fields on the base. Explicitly:
\begin{itemize}
\item We split into a base and fibre parts using the notation $X^{\u{\a}}=(\X^A,Y^{\u{a}})$ where the original (undoubled) co-ordinates $x^A$ of the $\X^A$ are compact and the original co-ordinates $y^a$ of the $Y^{\u{a}}$ are in general non-compact. This requires the reordering the indices so that the base co-ordinates and their duals sit beside each other in $Y^{\u{a}}$. We take the metric to have block-diagonal form
\beq
\H_{\u{\a}\u{\b}}= \left( \begin{array}{cc}
    \H_{AB}& 0 \\ 
    0 & \H_{\u{a}\u{b}} \\ 
  \end{array}\right)\, .
\eeq
\item We allow none of the fields to depend on the fibre co-ordinates, i.e. $\d_A\H_{AB}=\d_A\H_{\u{a}\u{b}}=0$.
\item We undouble the base co-ordinates. Similar to showing the equivalence of the double field theory to the ordinary string. Splitting $Y^{\u{a}}=(\tilde{y}_a,y^a)$ this amounts to setting $\dt^i=0$ on all fields. Once the base has been undoubled we will normally have the same index conventions as for the doubled formalism, with the splitting $X^\a=(\X^A,Y^a)$.
\item We take the anti-symmetric field $b$ to vanish on the base so that
\beq
H_{\u{a}\u{b}}=  \left(\begin{array}{cc}
    g^{ab} & 0 \\ 
    0 & g_{ab} \\ 
  \end{array}\right)\, .
\eeq
On the base $L_{\a\b}$ takes the form
\beq
L_{\u{a}\u{b}}=  \left(\begin{array}{cc}
    0 & \delta^{a}_{\ b} \\ 
    \delta_{a}^{\ b} & 0 \\ 
  \end{array}\right).
  \eeq
  \item The doubled dilaton is defined through
\beq
e^{-2d}=\sqrt{h}e^{-2\phi},
\eeq
where $h$ is the determinant of the undoubled metric on the whole space. From the block-diagonal form of $h'$  we deduce
\beq\label{Dvsphi}
\d_A d=-\frac{1}{4}g^{ab}\d_A g_{ab}+\d_A\Phi\, ,
\eeq
where $\Phi$ is the doubled dilaton for the doubled theory on the fibre only, defined through
\beq
e^{-2\Phi}=\sqrt{h'}e^{-2\phi},
\eeq
where this time $h'$ is the determinant of the undoubled metric on the fibre only. 
\end{itemize}

\subsection{Evaluating $K$}

There are 8 possible terms in $K_{\m\n}$ \footnote{We allow Greek indices to be indices on the doubled total space (which we before denoted with underlined Greek indices) for this subsection only to keep expressions clearer,  and use $\m,\n$ rather than $\a,\b$ so expressions can be identified with those of \cite{Hohm:2010pp}. Elsewhere as in \cite{Berman:2007xn} $\m$ and $\n$ will be reserved for worldsheet indices, but this should be obvious from context. }  without the dilaton. Taking each of these in turn we get
\bea
\frac{1}{8}\, \partial_{\m}{\cal H}^{\k\l}
  \,\partial_{\n}{\cal H}_{\k\l}&=&\frac{1}{8}\d_m\H^{KL}\d_n\H_{KL}+\frac{1}{4}\d_m g^{kl}\d_n g_{kl}\, ,\label{firstline}\\
  -{1\over 4} \partial_\l{\cal H}^{\l\k} \partial_\k {\cal H}_{\m\n}&=&-\frac{1}{4}\d_l g^{kl}\d_k\H_{MN}-\frac{1}{4}\d_l g^{kl}\d_k g_{mn}-\frac{1}{4}\d_l g^{kl}\d_k g^{mn}\, ,\\
  -{1\over 4} {\cal H}^{\l\k} \partial_\l\partial_\k {\cal H}_{\m\n}&=&-\frac{1}{4}\d^2\H_{MN}-\frac{1}{4}\d^2 g_{mn}-\frac{1}{4}\d^2 g^{mn}\, ,\\
-\frac{1}{2} \partial_{(\m}{\cal H}^{\k\l}\,\partial_{\l} {\cal H}_{\n)\k}&=&-\frac{1}{2}\d_{(m} g^{kl}\d_l g_{n)k}\, ,\\
\frac{1}{2}\d_\l\H^{\k\l}\d_{(\m}\H_{\n)\k}&=&\frac{1}{2}\d_l g^{kl}\d_{(m} g_{n)k}\, ,\\
\frac{1}{2}\H^{\k\l}\d_\l\d_{(\m}\H_{\n)\k}&=&\frac{1}{2}g^{kl}\d_l\d_{(m} g_{n)k}\, ,\\
\frac{1}{2}\d_\l\H^{\k}_{\ (\m}\d_\k\H^{\l}_{\ \n)}&=&\frac{1}{2}\d_l g^{k(m}\d_k g^{ln)}\, ,\\
\frac{1}{2}\H^{\k}_{\ (\m}\d_\l\d_\k\H^{\l}_{\ \n)}&=&\frac{1}{2}g^{k(m}\d_l\d_k g^{ln)}\, ,
\eea
where $(m\ldots n)$ indicates symmetrisation over only the two outermost indices, defined such that in is idempotent (i.e. $((m\ldots n))=(m\ldots n)$). Note that while doubled indices such as $\m,\n,\k,\l,\ldots$, $M,N,\ldots$ or $\u{m},\u{n}\dots$ are raised with the $O(p,p)$ metric $L$ for the relevant $p$, the undoubled indices $m,n,\ldots$ are raised with $g^{mn}$. It is important to notice that, for example, there are two contributions to the second term of (\ref{firstline}) because the base is doubled, that is
\beq
\frac{1}{8}\d_m \H^{kl}\d_n \H_{kl}=\frac{1}{8}\d_m g^{kl}\d_n g_{kl}+\frac{1}{8}\d_m g_{kl}\d_n g^{kl}=\frac{1}{4}\d_m g^{kl}\d_n g_{kl}\, .
\eeq

There are four terms in $\K_{\m\n}$ featuring the doubled dilaton $d$. Even when doubled formalism dilaton $\Phi=0$ these give which contributions to $\K_{\m\n}$ via the first term in (\ref{Dvsphi}). The four terms give us respectively
\bea
 {1\over 2} \partial_\l d{\cal H}^{\l\k} \partial_\k {\cal H}_{\m\n}&=&-\frac{1}{8}g^{ab}\d^k g_{ab}\d_k H_{MN}\, ,\\
2\d_\m\d_\n d&=&-\d_n g^{kl}\d_m g_{kl}-g^{kl}\d_n \d_m g_{kl}\, ,\\
-\d_\l d\H^{\k\l}\d_{(\m}\H_{\n)\k}&=&g^{kp}\d^qg_{kp}\d_{(m}g_{n)q}\, ,\\
-\d_\l d\H^{\k}_{\ (\m}\d_\k\H^{\l}_{\ \n)}&=&-\frac{1}{2}g^{kp}\d_lg_{kp}\d^lg_{mn}\,
\eea
(we will return to the case of non-zero $\Phi$ later). We note that these dilaton terms can be written as 
\beq
-\frac{1}{2}g^{kl}\d^p g_{kl}\left(\hat{D}_\mu(g_{pn}\d^\mu X^n)-\frac{1}{2}\d_p\H_{MN}\d_1 X^M\d_1 X^N\right)
\eeq
which is proportional to the equation of motion and so vanishes on-shell.

\subsection{Evaluating $\R$}

Recall that the generalised Ricci curvature of \cite{Hohm:2010pp} is actually given by 
\beq
\R_{\m\n}=\frac{1}{2}\left(K_{\m\n}-\H^{\ \k}_\m K_{\k\l}\H^{\l}_{\ \n}\right)\, .
\eeq
Combining all the fibre pieces of $K$ we obtain
\beq
K_{MN}=-\frac{1}{4}\d^2\H_{MN}-\frac{1}{4}\d_l g^{kl}\d_k H_{MN}-\frac{1}{8}g^{ab}\d^k g_{ab}\d_k H_{MN}
\eeq
and we can thus use
\beq
\H(\d^2\H^{-1})\H=2(\d_a\H)\H^{-1}(\d^a\H)-\d^2\H
\eeq
and
\beq
\G{k}{a}{b}g^{ab}=-\d_l g^{kl}-\frac{1}{2}g^{ab}\d^kg_{ab}
\eeq
to determine
\beq
\R_{MN}=-\frac{1}{4}\d^2\H_{MN}+\frac{1}{4}\left((\d_a\H)\H^{-1}(\d^a\H)\right)_{MN}+\frac{1}{4}\G{k}{a}{b}g^{ab}\d_k\H_{MN}.
\eeq
This we compare with 
\bea
-W_{MN}&=&-\frac{1}{2}\d^2\H_{MN}+\frac{1}{2}\left((\d_a\H)\H^{-1}(\d^a\H)\right)_{MN}+\frac{1}{2}\G{k}{a}{b}g^{ab}\d_k\H_{MN}\, .
\eea
Turning out attention to the base we observe
\beq
\H_{\u{m}}^{\ \u{k}}\K_{\u{k}\u{l}}\H^{\u{l}}_{\ \u{n}}=  \left(\begin{array}{cc}
    g^{mk}K_{kl}g^{ln}  &g^{mk}K_{k}^{\ l}g_{ln} \\ 
    g_{mk}K_{\ l}^{k}g^{ln} &g_{mk}K^{kl}g_{ln}  \\ 
  \end{array}\right).
\eeq
and since $K_{mn}$ has only diagonal pieces this leads to
\beq
\R_{\u{mn}}=  \frac{1}{2}\left(\begin{array}{cc}
    K^{mn}-g^{mk}K_{kl}g^{ln}  &0 \\ 
   0&K_{mn}-g_{mk}K^{kl}g_{ln}  \\ 
  \end{array}\right).
\eeq
Let us examine $\R_{mn}=(K_{mn}-g_{mk}K^{kl}g_{ln})/2$ noting that if it vanishes, then so will $\R^{mn}=-g^{mk}\R_{kl}g^{ln}$. So we must lower parts of $K$ with both base indices up and subtract them from those with both indices down. After using identities such as
\beq
g_{nl}\d_k\d_m g^{kl}=-\d_m g^{kl}\d_k g_{nl}-\d_k g^{kl}\d_m g_{nl}-\d^l\d_mg_{nl}
\eeq
we obtain the answer
\bea\label{plainRicci}
\R_{mn}&=&\d_k g^{kl}\d_m g_{ln}-\frac{1}{2}\d_k g^{kl}\d_l g_{mn}+\d^l\d_mg_{nl}-\frac{1}{2}\d^2g_{mn}\\
&&-\frac{1}{2}\d_n g^{kl}\d_m g_{kl}-\frac{1}{2}g^{kl}\d_n \d_m g_{kl}\\
&&\frac{1}{2}g^{kp}\d^qg_{kp}\d_mg_{qn}-\frac{1}{4}g^{kp}\d_lg_{kp}\d^lg_{mn}\\
&&+\frac{1}{2}\d^kg_{ml}g^{lq}\d_kg_{qn}-\frac{1}{2}\d^{k}g_{ml}\d^lg_{kn}-\frac{1}{4}g^{kp}g^{lq}\d_mg_{pl}\d_ng_{qk}\\
&&+\frac{1}{8}\d_m\H^{KL}\d_n\H_{KL}
\eea
where we recognise the Ricci tensor of $g$, $\hat{R}_{mn}$, and we conclude
\beq
2\R_{mn}=\hat{R}_{mn}+\frac{1}{8}\d_m\H^{KL}\d_n\H_{KL}=-W_{mn}.
\eeq
Clearly the vanishing of $\R_{\m\n}$ is equivalent to the vanishing of $W_{\m\n}$, with in fact  $\R_{\\m\n}=W'_{\m\n}\,$.
\subsection{Including the dilaton}

Working from the double field side lets us easily see what how the dilaton should fit into the beta-functionals, and agrees with the results found in \cite{Berman:2007yf}. Using (\ref{Dvsphi}) in the expressions for $\K_{\m\n}$ for non-zero $\Phi$ one finds additional contributions
\bea
\R_{\Phi\; MN}&=&\frac{1}{2}\d_l\Phi\d^l\H_{MN}\, ,\\
\R_{\Phi\; mn}&=&\d_m\d_n\Phi-\hat{\Gamma}^{l}_{\ mn}\d_l\Phi\, ,
\eea
which we recognise as
\bea
\R_{\Phi\; \m\n}=D_\m D_\n\Phi\, .
\eea
Given the results of the previous section, including this dilaton corresponds to a modification of the beta-functional of the form
\bea\label{Wndil}
W_{\m\n}\rightarrow W_{\m\n}-2D_\m D_\n\Phi\, ,
\eea
which is as expected from \cite{Berman:2007yf} (the factor of 2 is simply due to the different normalisation of the dilaton used there). Let us remind ourselves why contributing to the beta-functional in this way is a defining property of the dilaton\cite{Hull:1985rc}. Though vanishing of the beta-functional is sufficient for ultraviolet finiteness of the ordinary string sigma model, it is not necessary; if the Ricci tensor is of the form
\beq\label{Rdv}
R_{mn}=D_{(m}v_{n)}
\eeq
then at one-loop we can remove the divergence using a field redefinition. However, $R_{mn}=0$ is also required for the vanishing of the Weyl anomaly and hence the consistency of string theory in this background. What is more, a Ricci tensor of the form (\ref{Rdv}) gives a contribution to the Weyl anomaly that cannot be removed by a field redefinition. This contribution can however be removed by adding a finite local counterterm to the effective action if $v_n=D_n\Phi$ for some scalar $\Phi$. This counterterm is the Fradkin-Tseytlin term coupling the dilaton $\Phi$ to the string worldsheet. Hence in the presence of a dilaton the beta-functional of the metric is modified in the manner of (\ref{Wndil}). In \cite{Berman:2007yf} it was demonstrated that this generalised metric beta-functional including $\Phi$ reproduced the dilaton contributions to the beta-functionals of $g$ and $b$ in the undoubled sigma model (the first contribution to the dilaton beta-functional itself appears at two-loops in the background field method, but can be determined indirectly from the other beta-functionals via an integrability condition).


\section{Discussion}

We have shown the equivalence the vanishing of the one-loop beta-functional, or equivalently one-loop Weyl invariance, of the doubled formalism (expressed as the vanishing of $W$) and the vanishing of the generalised Ricci tensor in the doubled formalism, which follows from the Euler-Lagrange equation of the generalised metric $\K$. This was in a fibred background where the fields on the fibre did not depend on the fibre co-ordinates, which raises the question of whether there is a more general chiral sigma model which gives the generalised metric formulation of the doubled field theory without restricting to this fibred background. This is something we hope to report on in future work. There are of course difficulties in writing a more general sigma model for the doubled formalism, which is why the co-ordinate dependence of $\H$ was restricted in the first place, further work on this issue appears in \cite{Hull:2009sg,ReidEdwards:2009nu}. The utility of such a sigma model performing the background field expansion of such a sigma model to higher order would lead to higher-order corrections to the doubled field theory (this may be a more practical way of finding higher order corrections to the string theory effective action including $b$ automatically).

When we performed the background field expansion we did things in a different manner to\cite{Berman:2007xn}. We maintained the covariance of the expressions as much as possible, showing that it is not required to add terms proportional to the equations of motion. Adding these terms corresponds to a field redefinition and hence we showed such a redefinition is unnecessary. We also demonstrated that one is not able to drop `connection terms' on the fibre as these do contribute, unlike the standard string case, or indeed the fibre part of this calculation. Due to the chiral action on the fibre the connection no longer enters as a minimally coupled gauge field and arguments that it cannot contribute to the Weyl anomaly at this order no longer stand. Performing the calculation in this way is less involved and provides a better understanding that should also be useful in the search for a more general sigma model.  

It was remarked in \cite{Berman:2007xn} that $W_{MN}$ was not the Ricci tensor of $\H$, though it did take a similar form. In the more general case of the double field theory there are many more index structures that can contribute, and it is also clearer that $\R_{MN}$, the generalised Ricci tensor, is not the same as the Ricci tensor of $\H$. The connection of $\R_{MN}$ to the result of the background field expansion of a sigma model strengthens the case that it should be considered a Ricci tensor like object. A doubled geometrical picture in which this Ricci tensor emerges has also been sought recently in \cite{Jeon:2010rw,Hohm:2010xe} and in \cite{Jeon:2011cn} it was obtained from a construction involving a semi-covariant derivative whose connection depends on the doubled dilaton. Similar technology was used for a doubled Yang-Mills theory\cite{Jeon:2011kp}.

Another instance in which a Ricci like tensor occurs in a doubled theory is Poisson-Lie T-duality\cite{Klimcik:1995ux}. Indeed it follows after a background field expansion similar to those performed here\cite{Avramis:2009xi,Sfetsos:2009vt}. One can ask if this more group-theoretic object can also be related to the double field theory Ricci tensor we have discussed. More recently an M-theoretic equivalent of the doubled geometry has been proposed\cite{Berman:2010is,Berman:2011pe}. Here rather than extra dual co-ordinates for string winding modes you must introduce a greater number of dual co-ordinates representing possible membrane and fivebrane windings. This should be related to the double field theory by dimensional reduction\cite{Thompson:2011uw}. Then one can wonder about what new structures are needed for a differential geometry of this new generalised M-theory geometry and whether this might allow some insight beyond the supergravity approximation in M-theory. It the presence of fundamental extended objects (equivalently the form-fields they couple to) that leads to the new dual dimensions via winding (and the form-fields combine with the metric in a new geometric object). These new geometric objects make manifest hidden symmetries of the theory, and since the existence of these form-fields is linked to central charges from supersymmetry one wonders if things will become clearer if we are able to include fermions.

\section*{Acknowledgements}

I would very much like to thank David Berman and Dan Thompson for discussions and comments as well as Jeong-Hyuck Park for comments. This research is supported in part by the Belgian Federal Science Policy Office through the Interuniversity Attraction Pole IAP VI/11 and by FWO-Vlaanderen through project G011410N.


\pagestyle{plain}
\def\href#1#2{#2}
\bibliographystyle{BibliographyStyle}
\addcontentsline{toc}{chapter}{\sffamily\bfseries Bibliography}

\bibliography{DoubleBib}

\end{document}